\documentclass[showkeys,showpacs,prd]{revtex4}
\usepackage{graphicx}
\usepackage{amsmath}
\usepackage{amsfonts}
\usepackage{color}

\begin{document}

\title{SU(3) flux tube gluon condensate}
\author{Vladimir Dzhunushaliev}
\email{vdzhunus@krsu.edu.kg}
\affiliation{Dept. Theor. and Nucl. Phys., KazNU, Almaty, 010008, Kazakhstan; \\
Institut f\"ur Physik, Universit\"at Oldenburg, Postfach 2503
D-26111 Oldenburg, Germany
}

\begin{abstract}
The distribution of a gluon condensate in a flux tube is calculated. The result is that the chromoelectric fields are confined with a surrounding coset chromomagnetic field. Such picture presents the concrete realization of dual QCD model in a scalar model of the flux tube. In the scalar model the SU(3) gauge fields are separated on two parts: (1) is the $SU(2) \subset SU(3)$ subgroup, (2) is the coset $SU(3) / SU(2)$. The SU(2) degrees of freedom are almost classical and the coset degrees of freedom are quantum ones. A nonperturbative approach for the quantization of the coset degrees of freedom is applied. In this approach 2-point Green's function is a bilinear combination of scalar fields and 4-point Green's function is the product of 2-points Green's functions. The gluon condensate is an effective Lagrangian describing the SU(2) gauge field with broken gauge symmetry and coupling with the scalar field. Corresponding field equations give us the flux tube.
\end{abstract}


\maketitle

\section{Introduction}

The problem of quark confinement has been one of the central problem of the high energy physics for years, where according to the dual QCD picture the QCD vacuum behaves like a dual superconductor and the condensation of monopoles plays a vital role in the whole confinement mechanism. In the dual picture of QCD \cite{nambu}, the monopoles get condensed and an Abrikosov vortex is formed between quark - antiquark, which gives rise to the infinitely rising linear confining potential. For review of dual QCD, see \cite{baker}.

One of the problems in a nonperturbative QCD is the determination of the condensates of the QCD vacuum. It is well known that the condensates can only be determined in a nonperturbative formulation of the QCD. There is a long history of attempts to determine the gluon condensate from first principles \cite{Banks}.

The problem of the quantization of strong interaction is the problem of a nonperturbative quantization. The solution of this problem is connected with the search of nonperturbative technique. It means that in the nonperturbative quantization technique we \textcolor{blue}{\emph{must exclude the words quantum particles, propagators and so on}}, i.e. all words connected with the perturbative technique (Feynman diagrams).

In this paper we calculate the gluon condensate of flux tube and show that the dual QCD picture is realized in a scalar model of flux tube between quark - antiquark. In this model the SU(3) gauge potential
$A^B_\mu \in SU(3), B=1,2, \cdots , 8$ is separated on two parts:
\begin{itemize}
	\item the first one is the gauge components $A^b_\mu \in SU(2) \subset SU(3), b=1,2,3$ which is in a classical state;
	\item the second one is $A^m_\mu \in SU(3)/SU(2), m=4,5,6,7,8$ and it is in a quantum state.
\end{itemize}
One can say that $A^b_\mu$ is in an ordered phase and $A^m_\mu$ is in a disordered phase. The effective Lagrangian for such system (which up to a sign is the SU(3) gluon condensate) can be obtained using nonperturbative quantization method with some assumptions and approximations (see \cite{Dzhunushaliev:2010ab} for the application of this method for a glueball scalar model). Briefly saying in the scalar model the 2-point Green's function of color quantum field can be approximately presented as the product of scalar fields with some coefficients having color and Lorentzian indixes. 4-point Green's function is decomposed as the product of two 2-point Green's functions.

One can say that the quantum strongly interacting fields are not a cloud of quanta but is more similar to a turbulent fluid with a non-zero mean velocity and fluctuations about it. In QCD it corresponds to a longitudinal chromoelectric field in the flux tube and quantum fluctuations of the gauge field.

\section{Flux tube solution}

\begin{figure}[h]
\begin{minipage}[t]{.45\linewidth}
  \begin{center}
  \fbox{
  \includegraphics[width=.85\linewidth]{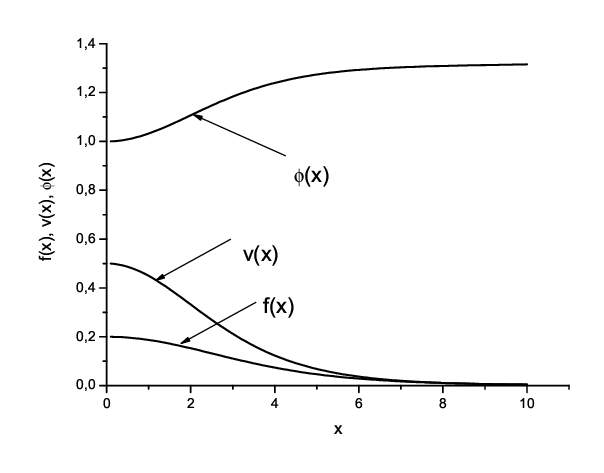}}
  \caption{The functions $f^*(x), v^*(x),\phi^*(x)$}
  \label{fig1}
  \end{center}
\end{minipage}\hfill
\begin{minipage}[t]{.45\linewidth}
  \begin{center}
  \fbox{
  \includegraphics[width=.85\linewidth]{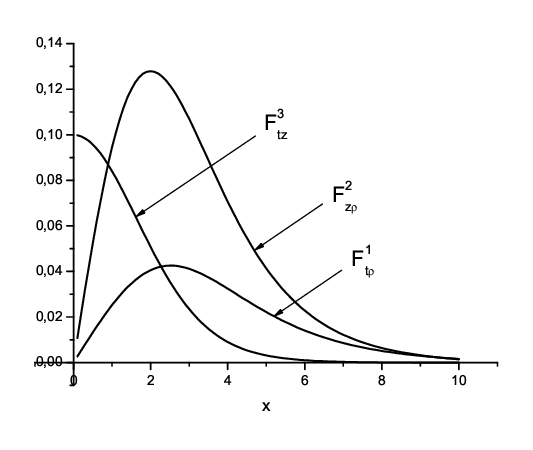}}
  \caption{The profiles of color fields
  $F^1_{t\rho}(x) = -f', F^2_{z\rho}(x) = -v', F^3_{tz}(x) = f(x) v(x)$.}
  \label{fig2}
  \end{center}
\end{minipage}\hfill
\end{figure}

In Appendix \ref{appendix} we have obtained an effective Lagrangian (which is up to a sign the SU(3) gluon condensate) describing the situation where there are ordered and disordered phases
\begin{equation}
\begin{split}
	&
	\mathcal L_{eff} = \left\langle \mathcal L_{SU(3)} \right\rangle =
	- \frac{1}{4 g^2}
	\left\langle
		\hat {\mathcal F}^{B\mu\nu} \hat {\mathcal F}^{B}_{\mu\nu}
	\right\rangle =
	- \frac{1}{4 g^2} F^{a}_{\mu\nu} F^{a \mu\nu} +
	\frac{1}{2} E^\mu_\nu \left| \phi_{,\mu} \right|^2 -
	\frac{\lambda}{4} \left(
		\left| \phi \right|^2 - \phi_\infty^2
	\right)^2 +
\\
	&
	\frac{1}{2} D^{ab \nu}_{\mu} A^a_\nu A^{b \mu}  \left| \phi \right|^2 -
	\frac{1}{2} \left( M^2 \right)^{ab \nu}_\mu A^a_\nu A^{b \mu} +
	\frac{1}{2} A^{a \nu} \left[
  	\left(B_1\right)^{a \mu}_\nu  \phi^* \phi_{,\mu}+
  	\left(B_2\right)^{a \mu}_\nu  \phi \phi^*_{,\mu}
  \right]
\label{1-10}
\end{split}
\end{equation}
where $F^a_{\mu \nu} = \partial_\mu A^a_\nu - \partial_\nu A^a_\mu + \epsilon^{abc} A^b_\mu A^c_\nu$ is the SU(2) field strength; $a,b,c = 1, 2,3$ are the SU(2) color indexes;
$\epsilon^{abc}$ are the structure constants for the SU(2) gauge group and $g$ is the coupling constant. We do not know the values of coefficients $E, D, M^2, B_{1,2}$ and for the simplicity we choose their in the form
\begin{equation}
\begin{split}
   E^\mu_\nu =& \delta^\mu_\nu,
\\
   D^{ab \nu}_{\mu} = & \delta^{ab} \delta^\nu_\mu,
\\
   \left( M^2 \right)^{ab \nu}_\mu = & m^2_a \delta^{ab} \delta^\nu_\mu,
   \text{ no summation over } a,
\\
   \left(B_1\right)^{a \mu}_\nu = & \left(B_2\right)^{a \mu}_\nu = 0 .
\end{split}
\label{1-20}
\end{equation}
Then the effective Lagrangian is
\begin{equation}
	\mathcal L_{eff} =
	- \frac{1}{4 g^2} F^{a}_{\mu\nu} F^{a \mu\nu} +
	\frac{1}{2} \left| \phi_{,\mu} \right|^2 -
	\frac{\lambda}{4} \left(
		\left| \phi \right|^2 - \phi_\infty^2
	\right)^2 +
	\frac{1}{2} A^a_\mu A^{a \mu}  \left| \phi \right|^2 -
	\frac{1}{2} m^2_a A^a_\nu A^{a \mu}
\label{1-25}
\end{equation}
The field equations are
\begin{eqnarray}
  \frac{1}{4g^2}D_\nu F^{a\mu\nu} &=& \left(
  	\left| \phi \right|^2 - m^2_a
  \right) A^{a \mu}, \text{ no summation over } a ,
\label{1-30}\\
  \phi_{,\mu}^{; \mu} &=& -\lambda \phi
  \left( \left| \phi \right|^2 - \phi_\infty^2
  \right) + A^a_\mu A^{a \mu} \phi .
\label{1-35}
\end{eqnarray}
The solution we search in the following form
\begin{equation}
    A^1_t(\rho) = f(\rho) ; \quad A^2_z(\rho) = v(\rho) ;
    \quad \phi(\rho) = \phi(\rho)
\label{1-40}
\end{equation}
here $z, \rho , \varphi$ are cylindrical coordinate system. The substitution
into equations \eqref{1-30} \eqref{1-35} gives us
\begin{eqnarray}
    f'' + \frac{f'}{x} &=& f \left( \phi^2 + v^2 - m^2_1 \right),
\label{1-50}\\
    v'' + \frac{v'}{x} &=& v \left( \phi^2 - f^2 - m^2_2 \right),
\label{1-60}\\
    \phi'' + \frac{\phi'}{x} &=& \phi \left[ - f^2 + v^2
    + \lambda \left( \phi^2 - \mu^2 \right)\right]
\label{1-70}
\end{eqnarray}
here we redefined $g \phi /\phi(0) \rightarrow \phi$, $f /\phi(0)  \rightarrow f$,
$v /\phi(0)  \rightarrow v$, $g \phi_\infty /\phi(0) \rightarrow \mu$,
$g m_{1,2} /\phi(0)  \rightarrow m_{1,2}$, $\rho \sqrt{\phi(0)}  \rightarrow x$. The color electric and magnetic fields are
\begin{equation}
  F^1_{t \rho} = -f', \quad F^2_{z \rho} = - v', \quad
  F^3_{tz} = fv .
\label{1-80}
\end{equation}
For the numerical calculations we choose the following parameters values
\begin{equation}
	\lambda = 0.1, \quad
	\phi (0) = 1, \quad
	v(0) = 0.5, \quad
	f(0) = 0.2.
\label{1-90}
\end{equation}
We apply the methods of step by step approximation for finding of numerical solutions (the details of similar calculations can be found in Ref. \cite{Dzhunushaliev:2003sq}). The numerical calculations give us the eigenvalues
$m_1^* \approx 1.23258$, $m_2^* \approx 1.18069$,
$\mu^* \approx 1.3136$
and eigenfunctions $v^*(x), f^*(x), \phi^*(x)$. The numerical solution is presented in Fig's. \ref{fig1} - \ref{fig2}. One can show that the flux of the chromoelectric field is nonzero
\begin{equation}
	\Phi = 2 \pi \int \limits_0^\infty \rho F^3_{tz} d \rho \neq 0.
\label{1-95}
\end{equation}

\section{The gluon condensate distribution in the flux tube}

\begin{figure}[h]
\begin{minipage}[t]{.45\linewidth}
  \begin{center}
  \fbox{
  \includegraphics[width=.85\linewidth]{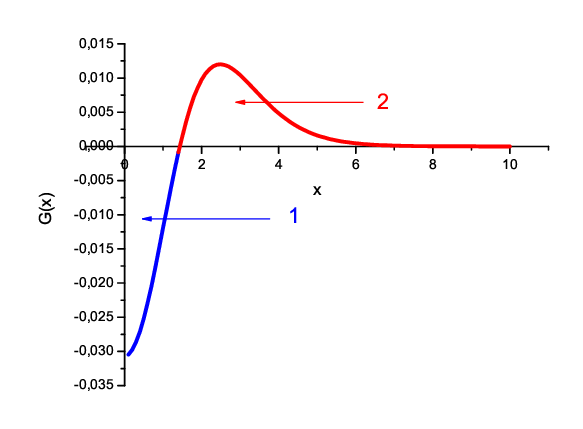}}
  \caption{The SU(3) gluon condensate $G(x)$. 1 is the area where chromoelectric fields are predominant. 2 is the area where chromomagnetic fields are predominant.}
  \label{fig3}
  \end{center}
\end{minipage}\hfill
\begin{minipage}[t]{.45\linewidth}
  \begin{center}
  \fbox{
  \includegraphics[width=.85\linewidth]{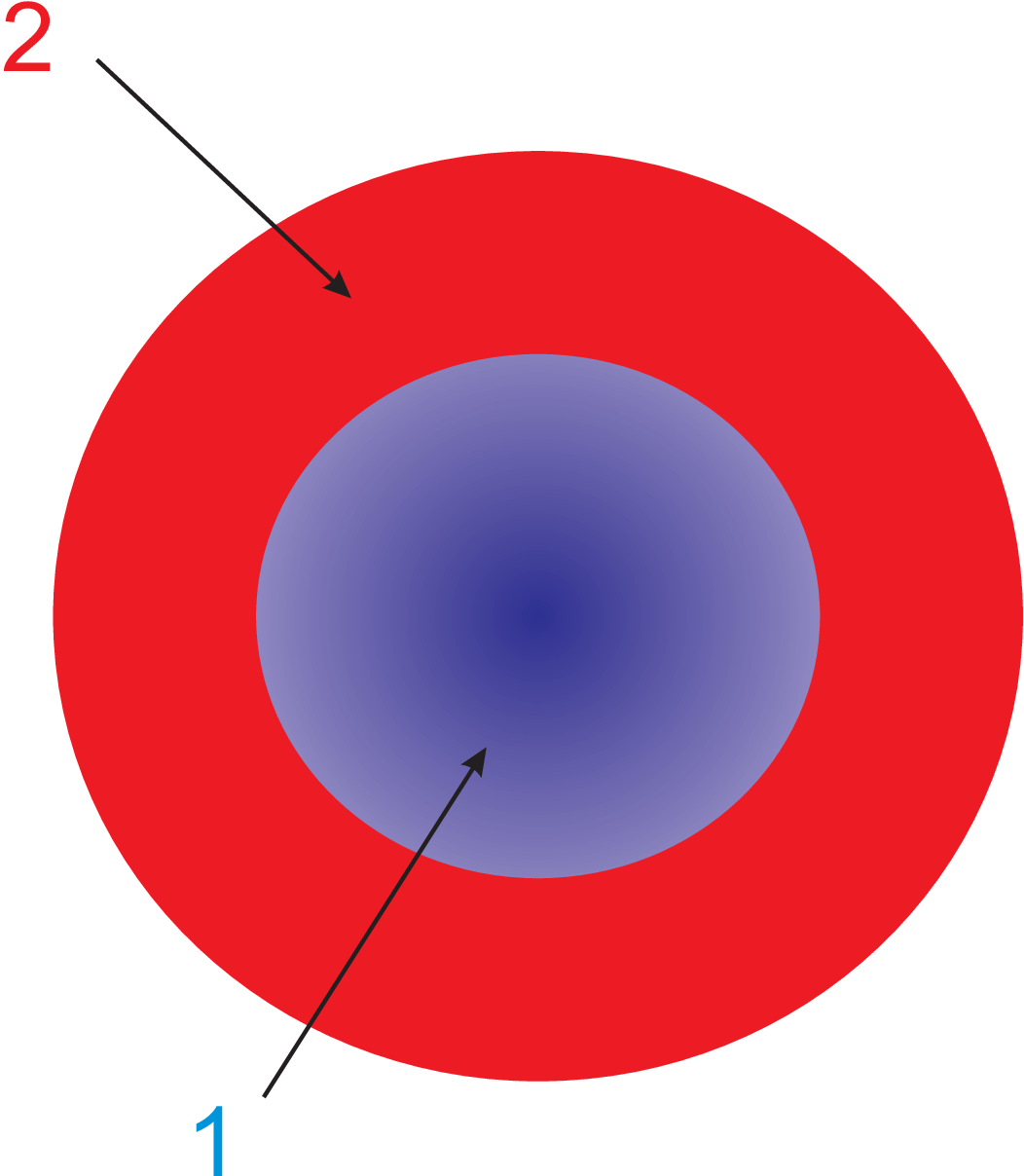}}
  \caption{Sketchy description of the distributions of chromoelectric and chromomagnetic fields: (1) \textcolor{blue}{blue} is the region where the chromoelectric field is predominant and (2) \textcolor{red}{red} is the region where the chromomagnetic field is predominant.}
  \label{fig4}
  \end{center}
\end{minipage}\hfill
\end{figure}

The distribution of the gluon condensate
$\left\langle \mathcal F^A_{\mu\nu} \mathcal F^{A\mu\nu} \right\rangle$ in the flux tube can be found from the effective Lagrangian \eqref{1-25}. The coset gluon condensate is
\begin{equation}
	G = - \mathcal L_{eff} =
	\left\langle \mathcal H^A_\mu \mathcal H^{A\mu} \right\rangle -
	\left\langle \mathcal E^A_\mu \mathcal E^{A\mu} \right\rangle
\label{2-10}
\end{equation}
where $\mathcal E^A_\mu, \mathcal H^A_\mu$ are chromoelectric and chromomagnetic fields. We see that if $G(x) < 0$ then in this area the chromoelectric field is predominant but if $G(x) > 0$ then the chromomagnetic field is predominant. The substitution of the ansatz \eqref{1-40} into the gluon condensate \eqref{2-10} gives us following
\begin{equation}
	G = - \frac{1}{2} {f'}^2 + \frac{1}{2} {v'}^2 - \frac{1}{2} f^2 v^2 +
	\frac{1}{2} m_1^2 f^2 - \frac{1}{2} m_2^2 v^2
	+ \frac{1}{2} {\phi'}^2 - \frac{1}{2} \left( f^2 - v^2 \right) \phi^2 +
	\frac{\lambda}{4} \left(
		\phi^2 - \mu^2
	\right)^2 .
\label{2-15}
\end{equation}
The profile of $G(x)$ is presented in Fig. \ref{fig3} and in Fig. \ref{fig4} one can see a schematical presentation of the flux tube with the distribution of chromoelectric and chromomagnetic fields. We see that the flux tube has a core (in \textcolor{blue}{blue}) where the quantum fluctuations of chromoelectric field and the longitudinal classical chromoelectric field $E^3_z$ are concentrated. These chromoelectric fields are confined by a belt (in \textcolor{red}{red}) filled with the chromomagnetic field. That is exactly what us say the dual QCD model:  In the dual superconductor picture of the QCD vacuum, chromomagnetic monopoles (creating chromomagnetic field) condense into dual Cooper pairs, causing chromoelectric flux to be squeezed into tubes. In this connection one can note that: (a) in Ref. \cite{Antonov:1998xt} it is derived the string representation of field correlators and it is shown that the obtained results are in agreement with the stochastic model of QCD vacuum; (b) it is possible that the gluon condensate is created not only by monopoles but dyons also \cite{Nandan:2004ms}.

\section{Discussion and conclusions}

Here we have shown that using nonperturbative quantization technique one can obtain a flux tube connecting quark - antiquark. In such tube a longitudinal chromoelectric field and quantum fluctuations of chromoelectric and chromomagnetic fields are presented. In our approximation the fluctuations of quantum gauge fields with a scalar field are described: 2-point Green's function is a bilinear combination of the scalar fields. The proportionality coefficients carry on themselves color and Lorentzian indexes. 4-point Green's function is the product of 2-point Green's functions. In this approximation one can average SU(3) Lagrangian and obtain an effective Lagrangian, i.e. to calculate gluon condensate. The effective Lagrangian describes SU(2) gauge field with broken gauge symmetry and coupling to the scalar field.

In the presented model the SU(3) gauge field is separated on ordered (almost classical $SU(2)$ gauge field) and disordered (quantum gauge fields belonging to the coset $SU(3) / SU(2)$) phases. The ordered phase is the longitudinal chromoelectric field. The disordered phase is the coset gluon condensate. We have found that the SU(3) gluon condensate in the flux tube has interesting space distribution: (a) at the core of the flux tube are the SU(2) longitudinal chromoelectric field and the quantum fluctuating coset chromoelectric field, (b) round the core is the quantum fluctuating coset chromomagnetic field. Such distribution confirms the dual QCD picture: the chromomagnetic field confines the chromoelectric fields into the tube.

Thus the correctness of the presented approach are completely borne out by two conclusions: (a) the flux tube solution is obtained, (b) the distribution of gluon condensate in flux tube is in agreement with the dual QCD picture.

\section*{Acknowledgments}

I am grateful to the Research Group Linkage Programme of the Alexander  von
Humboldt Foundation for the support of this research.

\appendix

\section{The calculation of gluon condensate - effective Lagrangian}
\label{appendix}

The SU(3) Lagrangian is
\begin{equation}
    \mathcal L_{SU(3)} = - \frac{1}{4 g^2} \mathcal F^{B\mu\nu} \mathcal F^{B}_{\mu\nu}
\label{1a-20}
\end{equation}
where $\mathcal F^B_{\mu \nu} = \partial_\mu A^B_\nu - \partial_\nu A^B_\mu + g f^{BCD} A^C_\mu A^D_\nu$
is the field strength; $B,C,D = 1, \ldots ,8$ are the SU(3) color indices; $a,b,c = 1,2,3$ are SU(2) color indices; $m,n,p = 4,5, \cdots , 8$ are coset indices; $g$ is the coupling constant; $f^{BCD}$ are the structure constants for the SU(3) gauge group and $g$ is the coupling constant.

Our goal is to average quantum version of the Lagrangian \eqref{1a-20} which is up to a sign the SU(3) gluon condensate
\begin{equation}
	\left\langle \mathcal L_{SU(3)} \right\rangle = - \frac{1}{4 g^2}
	\left\langle
		\hat {\mathcal F}^{B\mu\nu} \hat {\mathcal F}^{B}_{\mu\nu}
	\right\rangle
\label{1a-30}
\end{equation}
with some approximations concerning to 2 and 4-point Green's function
\begin{eqnarray}
  G^{mn}_{\mu \nu}(x_1, x_2) &=&
	\left\langle
		\hat A^m_\mu (x_1) \hat A^n_\nu (x_2)
	\right\rangle ,
\label{1a-40}\\
  G^{mnpq}_{\mu \nu \rho \sigma}(x_1, x_2, x_3, x_4) &=&
	\left\langle
		\hat A^m_\mu (x_1) \hat A^n_\nu (x_2) \hat A^p_\rho (x_3) \hat A^q_\sigma (x_4)
	\right\rangle
\label{1a-45}
\end{eqnarray}
where $\left\langle \cdots \right\rangle $ is a quantum averaging over some quantum state
$\left| \psi \right\rangle$, and $\hat A^m_\mu \in SU(3) / SU(2), m = 4,5, \cdots , 8$ in \eqref{1a-40} \eqref{1a-45} are the operators but $A^b_\mu \in SU(2), b=1,2,3$ are classical degrees of freedom. Below we will omit the operator symbol
$(\widehat {\cdots})$.

Taking into account that
\begin{eqnarray}	
	\mathcal F^{a}_{\mu\nu} &=& F^{a}_{\mu\nu} + f^{amn} A^m_\mu A^n_\nu ,
\label{1a-50}\\
	\mathcal F^{m}_{\mu\nu} &=& \left( F_0 \right)^{m}_{\mu\nu} +
	f^{mpq} A^p_\mu A^q_\nu + f^{mnb} \left(
		A^n_\mu A^b_\nu - A^n_\nu A^b_\mu
	\right)
\label{1a-60}
\end{eqnarray}
we have
\begin{equation}
\begin{split}
  \mathcal F^{B}_{\mu\nu} \mathcal F^{B \mu\nu} =
  &
  F^{a}_{\mu\nu} F^{a\mu\nu} +
  2 f^{amn} F^a_{\mu \nu} A^{m \mu} A^{n \nu} +
  f^{amn} f^{apq} A^m_\mu A^n_\nu A^{p \mu} A^{q \nu} +
\\
  &
  \left( F_0 \right)^{m}_{\mu\nu} \left( F_0 \right)^{m \mu\nu} +
  f^{mpq} f^{mrs} A^p_\mu A^q_\nu A^{r \mu} A^{s \nu} +
  f^{mna} f^{mpb} \left(
  	A^n_\mu A^a_\nu - A^n_\nu A^a_\mu
  \right) \left(
  	A^{p \mu} A^{b \nu} - A^{p \nu} A^{b \mu}
  \right) +
\\
  &
  f^{amn} \left[
  	\left( F_0 \right)^{m}_{\mu\nu} A^{p \mu} A^{q \nu} +
  	A^{p}_\mu A^{q}_\nu \left( F_0 \right)^{m \mu\nu}
  \right] +
\\
  &
  f^{mna} \left\{
  	\left( F_0 \right)^{m}_{\mu\nu} \left(
  		A^{n \mu} A^{a \nu} - A^{n \nu} A^{a \mu}
  	\right) + \left(
  		A^{n}_\mu A^{a}_\nu - A^{n}_\nu A^{a}_\mu
  	\right) \left( F_0 \right)^{m \mu\nu}
  \right\}
\\
  &
	f^{mpq} f^{mna} \left[
		A^p_\mu A^q_\nu \left(
			A^{n \mu} A^{a \nu} - A^{n \nu} A^{a \mu}
		\right) +
	\left(
		A^n_\mu A^a_\nu - A^n_\nu A^a_\mu
	\right) A^{p \mu} A^{q \nu}
	\right]
\label{1a-70}
\end{split}
\end{equation}
where
$F^a_{\mu \nu} = \partial_\mu A^a_\nu - \partial_\nu A^a_\mu + f^{abc} A^b_\mu A^c_\nu \in SU(2) \subset SU(3)$;
$\left( F_0 \right)^{m}_{\mu\nu} = \partial_\mu A^m_\nu - \partial_\nu A^m_\mu$. For the calculation of the gluon condensate we need some information about 2 and 4-points Green's functions. We will assume that in our approximation
\begin{eqnarray}
	\left( G_2 \right)^{mn}_{\mu \nu}(x_1, x_2) =
	\left\langle
		A^m_\mu (x_1) A^n_\nu (x_2)
	\right\rangle &\approx& C^{mn}_{\mu \nu} \phi(x_1) \phi^*(x_2) +
	\tilde m^{mn}_{\mu \nu},
\label{1a-80}\\
	\left\langle
		A^m_\mu (x_1) A^n_\nu (x_2) A^p_\rho (x_3) A^q_\sigma (x_4)
	\right\rangle &\approx&
	\left\langle
		A^m_\mu (x_1) A^n_\nu (x_2)
	\right\rangle
	\left\langle
		A^p_\rho (x_3) A^q_\sigma (x_4)
	\right\rangle
\label{1a-90}
\end{eqnarray}
with following properties
\begin{eqnarray}
	C^{mn}_{\mu \nu} &=& C^{mn}_{\nu \mu},
\label{1a-100}\\
	\tilde m^{mn}_{\mu \nu} &=&
	\tilde m^{mn}_{\nu \mu}.
\label{1a-110}
\end{eqnarray}
The ansatz \eqref{1a-80} \eqref{1a-90} is similar to a model in which the QCD vacuum is simulated by a stochastic background field \cite{Dosch:1987sk} \footnote{In this connection one can note that in Ref. \cite{Shoshi:2002rd} on the basis of the stochastic vacuum model the chromofield distributions of static color dipoles were calculated. There it was shown that the nonperturbative stochastic vacuum model leads to confinement of the color charges in the dipole via a string of color fields.}. The difference is that we use the ansatz \eqref{1a-80} \eqref{1a-90} for the calculation of an effective Lagrangian describing the scalar field describing 2-point Green's function.

Using \eqref{1a-80} one calculate
\begin{equation}
	\left\langle \left(
		\partial_\mu A^m_\nu(x) \right)  A^n_\alpha(x)
	\right\rangle = \left.
		\frac{\partial}{\partial y^\mu}
		\left\langle
			A^m_\nu(y) A^n_\alpha(x)
		\right\rangle
	\right|_{y \rightarrow x} =
	C^{mn}_{\nu \alpha} \partial_\mu \phi(x) \phi^*(x).
\label{1a-115}
\end{equation}
For the calculation of the gluon condensate \eqref{1a-30} we need to take into account the decomposition \eqref{1a-70} and the form for 2 and 4-point Green's functions \eqref{1a-80} \eqref{1a-90}. At first we will calculate
\begin{equation}
	F^a_{\mu \nu} A^{m \mu} A^{n \nu} = 0
\label{1a-120}
\end{equation}
as $F^a_{\mu \nu} = - F^a_{\nu \mu}$ is antisymmetric but
$\left( G_2 \right)^{mn}_{\mu \nu} = \left( G_2 \right)^{mn}_{\nu \mu}$ is symmetric. Using \eqref{1a-80} \eqref{1a-90} we have
\begin{equation}
	f^{amn} f^{apq} \left\langle
		A^m_\mu A^n_\nu A^{p \mu} A^{q \nu}
	\right\rangle = \lambda_1 \left(
		\left| \phi \right|^2 - \phi_\infty^2
	\right)^2 + C_1
\label{1a-130}
\end{equation}
where $\lambda_1 = f^{amn} f^{apq} C^{mn}_{\mu \nu} C^{pq \mu \nu}$;
$-2 \phi_\infty^2 \lambda_1 = f^{amn} f^{apq} \left[
C^{mn}_{\mu \nu}\tilde m^{pq \mu \nu} +
C^{pq \mu \nu}\tilde m^{mn}_{\mu \nu}
\right]$ and
$\phi_\infty^4 \lambda_1 + C_1 = f^{amn} f^{apq}
\tilde m^{mn}_{\mu \nu} \tilde m^{pq \mu \nu}$. The same for
\begin{equation}
	f^{mpq} f^{mrs} \left\langle
		A^p_\mu A^q_\nu A^{r \mu} A^{s \nu}
	\right\rangle = \lambda_2 \left(
		\left| \phi \right|^2 - \phi_\infty^2
	\right)^2 + C_2
\label{1a-140}
\end{equation}
where $\lambda_2 = f^{mpq} f^{mrs} C^{pq}_{\mu \nu} C^{rs \mu \nu}$;
$-2 \phi_\infty^2 \lambda_2 = f^{mpq} f^{mrs} \left[
C^{pq}_{\mu \nu}\tilde m^{rs \mu \nu} +
C^{rs \mu \nu}\tilde m^{pq}_{\mu \nu}
\right]$ and
$\phi_\infty^4 \lambda_2 + C_2 = f^{mpq} f^{mrs}
\tilde m^{pq}_{\mu \nu} \tilde m^{rs \mu \nu}$. Next is the calculation of
\begin{equation}
	\left\langle
		\left( F_0 \right)^{m}_{\mu\nu} \left( F_0 \right)^{m \mu\nu}
	\right\rangle	=
	2 C^{mm \mu}_{\mu} \left| \phi_{,\mu} \right|^2 -
	2 C^{mm \mu}_{\nu} \phi_{,\mu} \phi^{* \nu} =
	- 2 E^\mu_\nu \phi_{,\mu} \phi^{*\nu}.
\label{1a-150}
\end{equation}
The next is
\begin{equation}
\begin{split}
	&
	f^{mna} f^{mpb} \left\langle
		\left(
  	A^n_\mu A^a_\nu - A^n_\nu A^a_\mu
  \right) \left(
  	A^{p \mu} A^{b \nu} - A^{p \nu} A^{b \mu}
  \right)
	\right\rangle	=
	\left(
		\tilde C^{ab} \delta^\nu_\mu - \tilde C^{ab \nu}_\mu
	\right) A^a_\nu A^{b \mu} \left| \phi \right|^2 +
\\
	&
	\left[
	m^{ab} \delta^\nu_\mu -
	m^{ab \nu}_\mu
	\right] A^a_\nu A^{b \mu} =
	- 2 D^{ab \nu}_{\mu} A^a_\nu A^{b \mu}  \left| \phi \right|^2 +
	2 \left( M^2 \right)^{ab \nu}_\mu A^a_\nu A^{b \mu}
\label{1a-160}
\end{split}
\end{equation}
where $\tilde C^{ab} = 2 f^{mna} f^{mpb} C^{np \mu}_\mu$;
$\tilde C^{ab \nu}_\mu = 2 f^{mna} f^{mpb} C^{np \nu}_\mu$;
$m^{ab} = 2 f^{mna} f^{mpb} \tilde m^{np \mu}_{\mu}$ and
$m^{ab \nu}_\mu = 2 f^{mna} f^{mpb} \tilde m^{np \nu}_{\mu}$. The next is
\begin{equation}
\begin{split}
	&
	f^{mna} \left\langle
  	\left( F_0 \right)^{m}_{\mu\nu} \left(
  		A^{n \mu} A^{a \nu} - A^{n \nu} A^{a \mu}
  	\right) + \left(
  		A^{n}_\mu A^{a}_\nu - A^{n}_\nu A^{a}_\mu
  	\right) \left( F_0 \right)^{m \mu\nu}
  \right\rangle =
\\
  &
  2 f^{mna} A^{a \nu} \left[ \left(
  	C^{mn \mu}_\nu - C^{mn \alpha}_\alpha \delta^\mu_\nu
  	\right) \phi^* \phi_{,\mu} +
  	\left(
  	C^{nm \mu}_\nu - C^{nm \alpha}_\alpha \delta^\mu_\nu
  	\right) \phi \phi^*_{,\mu}
  \right] =
  2 A^{a \nu} \left[
  	\left( B_1 \right)^{a \mu}_\nu  \phi^* \phi_{,\mu} +
  	\left(B_2 \right)^{a \mu}_\nu  \phi \phi^*_{,\mu}
  \right]
\label{1a-170}
\end{split}
\end{equation}
The next is
\begin{equation}
	\left\langle
		A^p_\mu A^q_\nu \left(
			A^{n \mu} A^{a \nu} - A^{n \nu} A^{a \mu}
		\right) +
	\left(
		A^n_\mu A^a_\nu - A^n_\nu A^a_\mu
	\right) A^{p \mu} A^{q \nu}
	\right\rangle = 0
\label{1a-180}
\end{equation}
as we assume as usually that schematically $\left( A^m_{\cdots} \right)^3 = 0$. Collecting all together we have
\begin{equation}
\begin{split}
	&
	\left\langle \mathcal L_{SU(3)} \right\rangle = - \frac{1}{4 g^2}
	\left\langle
		\hat {\mathcal F}^{B\mu\nu} \hat {\mathcal F}^{B}_{\mu\nu}
	\right\rangle =
	- \frac{1}{4 g^2} F^{a}_{\mu\nu} F^{a \mu\nu} +
	\frac{1}{2} E^\mu_\nu \phi_{,\mu} \phi^{*\nu} -
	\frac{\lambda}{4} \left(
		\left| \phi \right|^2 - \phi_\infty^2
	\right)^2 +
\\
	&
	\frac{1}{2} D^{ab \nu}_{\mu} A^a_\nu A^{b \mu}  \left| \phi \right|^2 -
	\frac{1}{2} \left( M^2 \right)^{ab \nu}_\mu A^a_\nu A^{b \mu} +
	\frac{1}{2} A^{a \nu} \left[
  	\left(B_1\right)^{a \mu}_\nu  \phi^* \phi_{,\mu} +
  	\left(B_2\right)^{a \mu}_\nu  \phi \phi^*_{,\mu}
  \right]
\label{1a-190}
\end{split}
\end{equation}
here we have redefined $\phi/g \rightarrow \phi$ and
$g^2(\lambda_1 + \lambda_2) \rightarrow \lambda$. All coefficients
$E, \lambda, \phi_\infty, D, M^2, B_{1,2}$ are linear or bilinear combinations of $C^{\cdots}_{\cdots}$ and $\tilde m^{\cdots}_{\cdots}$ and we omit unessential constant. Finally we have the Lagrangian describing SU(2) gauge theory coupling to a scalar field. Let us note that the sum
\begin{equation}
	\frac{1}{2} E^\mu_\nu \phi_\mu \phi^{*\nu} +
	\frac{1}{2} A^{a \nu} \left[
  	\left(B_1\right)^{a \mu}_\nu  \phi^* \phi_{,\mu} +
  	\left(B_2\right)^{a \mu}_\nu  \phi \phi^*_{,\mu}
  \right]	+
	\frac{1}{2} D^{ab \nu}_{\mu} A^a_\nu A^{b \mu}  \left| \phi \right|^2
\label{1a-200}
\end{equation}
is similar to the kinetic term for a color scalar field in corresponding Lagrangian
\begin{equation}
	D_\mu \phi^a D^\mu {\phi^*}^a =
	\left|
		\partial_\mu \phi^a + \epsilon^{abc} A^b_\mu \phi^c
	\right|^2 .
\label{1a-220}
\end{equation}

\end{document}